# Crack propagation in quasi-brittle materials by fourth-order phase-field cohesive zone model

Khuong D. Nguyen[a,b,c], Cuong-Le Thanh[d], Frank Vogel[e], H. Nguyen-Xuan[f*], M. Abdel-Wahab[f,h*]

[a]*Department of Electrical Energy, Metals, Mechanical Constructions & Systems, Faculty of Engineering and Architecture, Ghent University, Belgium*
[b]*Department of Engineering Mechanics, Faculty of Applied Science, Ho Chi Minh City University of Technology (HCMUT), Ho Chi Minh City, Vietnam*
[c]*Vietnam National University, Ho Chi Minh City, Vietnam*
[d]*Faculty of Civil Engineering, Ho Chi Minh City Open University, Ho Chi Minh City, Vietnam*
[e]*InuTech GmbH, 90429 Nuernberg, Germany*
[f]*CIRTech Institute, Ho Chi Minh City University of Technology (HUTECH), Ho Chi Minh City, Vietnam*
[h]*Soete Laboratory, Faculty of Engineering and Architecture, Ghent University, Belgium*

* Corresponding author
*Email addresses:* ngx.hung@hutech.edu.vn (H. Nguyen-Xuan); magd.a.w@hutech.edu.vn; magd.abdelwahab@ugent.be (M. Abdel-Wahab)

**Abstract**

A phase-field approach becomes a more popular candidate in modeling crack propagation. It uses a scalar auxiliary variable, namely a phase-field variable, to model a discontinuity zone in a continuity domain. Furthermore, the fourth-order phase-field approach produces a better convergence rate and more accuracy of the solutions than the second-order one. However, it is available for modeling the crack propagation in the brittle material. This study addresses the fourth-order phase-field model combining the non-standard phase-field form with a cohesive zone model (CZM) to predict crack propagation in the quasi-brittle material. A Cornelisson's softening law is used to capture the high precision of crack propagation prediction. The concrete material is considered as a quasi-brittle one. For computation efficiency using NURBS-based finite elements, Virtual Uncommon-Knot-Inserted Master-Slave (VUKIMS) technique is employed to derive a local refinement mesh. Numerical results are verified by the published ones. It was found that the peak load and crack path results are independent of the element size and insensitive to the length-scale number using the fourth-order phase-field CZM. The findings show the most significant advantage compared to the standard phase-field approach in terms of computational cost and solution accuracy.

1. **Introduction**

Fracture mechanics is a crucial area to understand crack initiation and growth in solid mechanics. The study concentrates on computing the failure behaviors of the structures which are subjected to multiple loading types in multi-physics problems, for instance, mechanical force [1], thermal [2], electro [3], magneto [4], and hydraulic fracturing [5]. In the earliest works, some exemplary methods have been proposed to obtain an accurate prediction of fracture mechanisms by the stress intensity factor (SIF) concept, e.g., Griffith [6], Irwin [7], Francfort [8], Buliga [9], Dal Maso [10], and Bourdin [11]. SIF is used to evaluate the energy of the crack tip zone. Several numerical methods have been developed to compute cracks as discontinuous zones in a continuous domain with the linear elastic fracture mechanics (LEFM) framework. First of all, the standard finite element method (FEM) used the re-meshing technique [12] to implement crack propagation. However, this technique requires significant time for generating a new mesh when the crack propagates. This issue is further aggravated in the three-dimensional problem. Next, through the partition of unity method (PUM), the enriched formulation family [1, 13-15] have been developed to circumvent the limitation of the re-meshing strategies by using a discontinuous interpolation of the displacement variable within enriched elements. However, these approaches have difficulties in computing the crack behavior in initiation, merging and branching, and even three-dimensional fracture problems.

Moreover, the task of crack propagation prediction is still a difficult challenge for the quasi-brittle material in the case of multi-cracks problems. An FEM-based approach is still a popular choice to compute the failure of the quasi-brittle structures. Some progressive approaches are investigated in this scope, for instance, cohesive zone model (CZM) [16, 17], strong discontinuity approach (SDA) [18, 19], and XFEM [20, 21]. These methods implement the crack as a discontinuity zone by creating traction-free crack surfaces or discontinuity of displacement. Also, in the fracture zone of quasi-brittle material, softening curves are established to determine a relationship between the crack opening width and the transferred cohesive stress. When more complex crack problems are considered, some drawbacks occur from the explicit representation of cracks. Because this description requires a re-meshing procedure and crack tracking approach with particular criteria for crack path prediction, the above approaches will have several difficulties of crack branching and merging issues in 3D.

During the past decade, a phase-field model, first proposed by Miehe [22], has recently attracted much attention from scientists in fracture mechanics. A phase-field model is a promising approach because it helps overcome the difficulties of the aforementioned approaches, including crack initiation, merging, branching, kinking, and nucleation [23-26], even in three-dimensional fracture [24, 27, 28]. The phase-field approach models the crack patterns under a scalar auxiliary variable, namely a *phase-field* variable. Crack propagation is tracked automatically by this variable without any criterion. Thus, a strong coupling problem between the phase-field and displacement fields is considered for this approach. In addition, it has been developed successfully on several complicated fracture problems, for instance, polycrystalline materials [29], piezoelectric and ferroelectric materials [30], thermo-elastic solids [31], multi-scale problems [32], microstructures [33], fluid-saturated porous media [34], and fatigue problems [35]. However, these studies only focus on brittle [24, 36-39] and ductile fracture [28, 40-44]. A few studies have been developed for the quasi-brittle material with the standard phase field formulation under cohesive fracture [45-49]. However, they have been applied to elementary geometries. Also, both sharp crack topology and a peak load of these numerical results are still dependent on the length-scale number $l_0$ and unstable with the element size. Furthermore, the general softening laws of general quasi-brittle fracture are not considered in these studies.

Recently, the unified phase-field damage model, as a non-standard phase-field model, has been introduced by Wu [50] to overcome the above issues of the standard one. It can be applied to both brittle and quasi-brittle material structures. The general softening laws, including linear, exponential, hyperbolic, Cornelisson's softening laws, are integrated into the approach to approximate the quasi-brittle material behaviors with high precision. Furthermore, Wu has demonstrated in the literature that the proposed approach is both insensitive to the length-scale number [51] and independent of the element size [52] through optimal characteristic functions. These characteristics cannot be reached from the standard phase-field models corresponding to each softening law. However, the non-standard phase-field formulation is still under the second-order phase-field model containing the first derivative of the phase-field variable. Regarding this point, Borden [36] developed a fourth-order phase-field formulation for brittle failure extending the original phase-field model [22]. The presence of higher-order derivatives of the phase-field variable has a higher convergence rate and a more accurate solution than the second-order one. However, the higher-order derivatives in the fourth-order one require the basis functions that fulfill at least $C^1$-continuity, which is difficult to satisfy using the traditional FEM. Isogeometric analysis (IGA) [53] shows the best candidate in this field. Its approximation is constructed by the Non-uniform rational B-Splines (NURBS) basis function. The interpolation order of basis functions is chosen very flexibly to determine the degree of smoothness. Thus, it makes IGA becomes a practical computational approach for the phase-field approach using the fourth-order formulation. However, the computational cost is a big complaint because the approach requires a small enough element size to describe a crack topology. Typically, an effective element size of $h$, recommended by several pieces of literature [38, 54-56] for the traditional FEM, is five times smaller than the length-scale parameter $l_0$. However, we proved that accurate solutions could be obtained using a cubic NURBS element with an effective element size $h = l_0 / 2$ [57].

In addition, the traditional IGA uses a tensor product NURBS basis function which causes a large number of wasted elements for computing due to using a global refinement mesh. Therefore, a local refinement technique, namely Virtual Uncommon-Knot-Inserted Master-Slave (VUKIMS) [58], is introduced to predict crack propagation in the brittle material from the previous study [57]. This technique helps to reduce both required memory and computational cost. Another technique, proposed by Rabczuk [59], uses hierarchical T-meshes (PHT-splines) to build a local refinement mesh. However, these studies only focused on simulating crack propagation of the brittle failure. Investigating the quasi-brittle fracture has not been addressed in the literature yet.

In this paper, we investigate the fourth-order phase-field formulation combining a non-standard phase-field model for predicting crack propagation of quasi-brittle failure. NURBS-based finite element approximations are employed to satisfy at least $C^1$-continuity, which is a requirement of the fourth-order formulation. Furthermore, the VUKIMS algorithm is also integrated into the approach to decrease the degrees of freedom in computations. After describing a fundamental formulation of the fourth-order phase-field CZM, Cornelissen's softening curve is considered to simulate concrete behaviors by the optimal coefficients of the energetic degradation function. In Section 2.6, we establish a generalized formulation available for both brittle and quasi-brittle materials. Section 3 shows several examples that have been considered to verify and demonstrate the accuracy of the solutions. Also, we demonstrated that the present results are insensitive to the length-scale number and independent of the element size. These factors overcome the shortcomings of the standard phase-field model. Finally, Section 4 summarizes the main points showing that the proposed method is a promising approach for a phase-field model for both brittle and quasi-brittle materials.

## 2. A brief review of the fourth-order phase-field theory approach

This section introduces the development of the fourth-order formulation combined with the cohesive zone model (CZM) to compute crack propagation in concrete material as the quasi-brittle material. This approach is developed from Wu's study [50], which used the second-order one.

### 2.1. Crack surface density functions

Francfort [8] and Bourdin [60] firstly investigated a phase-field approach for brittle failure based on a phase-field scalar number ($\phi$). The phase-field scalar number, $\phi \in [0,1]$, is considered to illustrate a smeared crack (please see Figure 1). For instance, $\phi = 1$ represents the discontinuity (fractured) domain while $\phi = 0$ depicts the remaining domain. Wu [50] has recently introduced a non-standard phase-field model to predict crack propagation in quasi-brittle material by combining energetic degradation functions and genetic geometric. The total energy function, including fracture energy and elastic energy, is defined as follows:

$$\Pi(\boldsymbol{\varepsilon},\phi) = \int_\Omega (g(\phi)\psi_e^+(\boldsymbol{\varepsilon}) + \psi_e^-(\boldsymbol{\varepsilon}))\mathrm{d}\Omega + \int_\Omega \mathcal{G}_C \psi_{\phi,n} \mathrm{d}\Omega \qquad (1)$$

where $g(\phi)$ is an energetic degradation function, $\mathcal{G}_C$ is the critical energy release density, and $\psi_{\phi,n}$ is a crack surface density function where $n$ corresponds to the use of the order phase-field model. The small strain tensor $\boldsymbol{\varepsilon}$ can be defined as $\boldsymbol{\varepsilon} = \mathrm{symm}[\nabla \mathbf{u}]$. Damage is considered on tension parts through the energetic degradation function ($g(\phi)$) defined in the following section. Eq. (1) can split the energy into tension and compression parts by using the positive ($\psi_e^+$) and negative ($\psi_e^-$) components, which are given as

$$\psi_e^{\pm}(\boldsymbol{\varepsilon}) = \frac{\lambda}{2}(\langle\mathrm{tr}(\boldsymbol{\varepsilon})\rangle^{\pm})^2 + \mu\mathrm{tr}[(\boldsymbol{\varepsilon}^{\pm})^2] \qquad (2)$$

where $\lambda$ and $\mu$ are Lamé's first parameter and shear modulus, respectively.

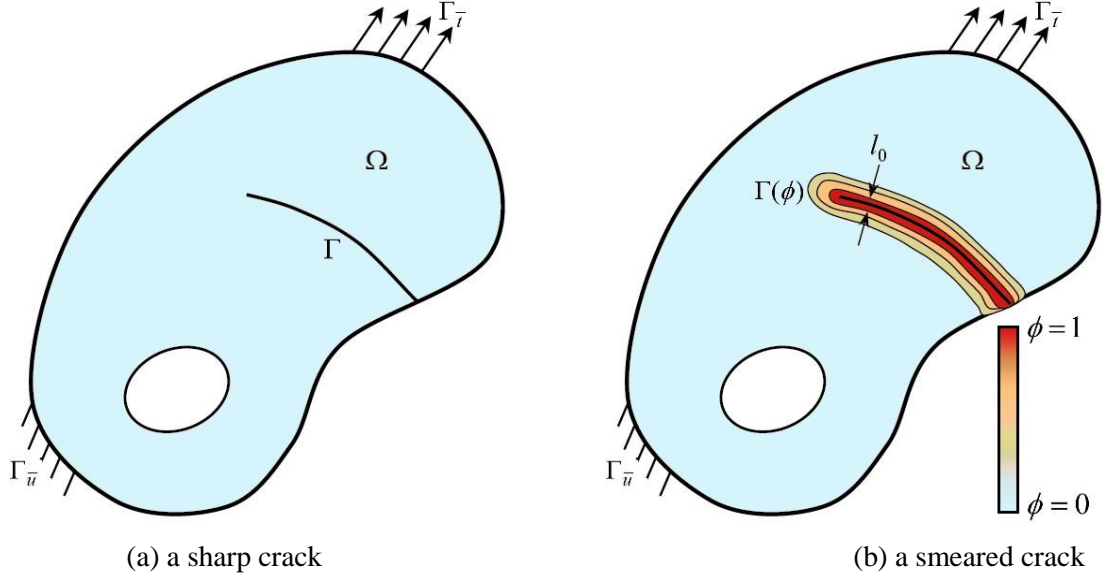

(a) a sharp crack  (b) a smeared crack

Figure 1. An internal crack can be represented by two approaches.

Positive and negative strain tensors are calculated from the strain tensor, as given:

$$\boldsymbol{\varepsilon} = \boldsymbol{\varepsilon}^+ + \boldsymbol{\varepsilon}^- \quad \text{with} \quad \boldsymbol{\varepsilon}^{\pm} = \sum_{i=1}^{d} \langle \varepsilon_i \rangle^{\pm} \mathbf{n}_i \otimes \mathbf{n}_i \qquad (3)$$

where $\varepsilon_i$ and $\mathbf{n}_i$ are the eigenvalues and eigenvectors of the strain tensor $\boldsymbol{\varepsilon}$, respectively. Macaulay brackets are defined as $\langle \cdot \rangle^{\pm} = \frac{1}{2}(\cdot \pm |\cdot|)$, as shown in Eqs. (2) and (3).

A crack surface density function of second-order theory for quasi-brittle fracture, which was proposed by Wu [50], is expressed as

$$\psi_{\phi,2} = \frac{1}{c_\alpha}\left(\frac{\alpha(\phi)}{l_0} + l_0 (\nabla\phi)^2\right) \qquad (4)$$

where $c_\alpha$ is a scaling number defined as $c_\alpha = 4\int_0^1 \sqrt{\alpha(\beta)}\mathrm{d}\beta$, and $\alpha(\phi) = \chi\phi + (1-\chi)\phi^2$ is the geometric crack function in the quadratic form with the non-negative scalar $\chi \in [0,2]$. A length-scale number, $l_0$, represents the width of the crack surface. Significantly, the geometric function becomes $\alpha(\phi) = \phi^2$ when $\chi = 0$ is chosen, Eq. (4) transforms to traditional crack geometric function which is given by Miehe [22]. In addition, Borden [36] has proposed a fourth-order formulation for the brittle material. He also proved that the numerical solutions, using the fourth-order approach, can be obtained with a higher convergence rate and a better accuracy than the second-order one in his study.

The fourth-order formulation was used successfully in our previous study [57] for crack propagation in the brittle material using NURBS-based finite element analysis. This study first develops the fourth-order phase-field CZM on a modified crack surface density function to compute crack propagation in the quasi-brittle material. The crack surface density function is determined as

$$\psi_{\phi,4} = \frac{1}{c_\alpha}\left(\frac{\alpha(\phi)}{l_0} + \frac{l_0}{2}(\nabla\phi)^2 + \frac{l_0^3}{16}(\Delta\phi)^2\right). \tag{5}$$

*2.2. Governing balance equations*

A coupled field problem is proposed to compute in fracture problem using a phase-field model. This approach includes displacement and phase-field variables. The strong form of the governing equation is expressed for static solid mechanics problem with the displacement field as follows:

$$\nabla\cdot\boldsymbol{\sigma} + \mathbf{b} = 0 \text{ on } \Omega \tag{6}$$

with the Neumann-type boundary conditions $\boldsymbol{\sigma}\cdot\mathbf{n} = \mathbf{t}$ on $\partial\Gamma_t$, where $\mathbf{t}$ is the traction force on the boundary $\partial\Gamma_t$ and $\mathbf{b}$ is the body force. The stress tensor, containing both positive and negative components, is given:

$$\boldsymbol{\sigma} = g(\phi)\frac{\partial\psi_e^+}{\partial\boldsymbol{\varepsilon}} + \frac{\partial\psi_e^-}{\partial\boldsymbol{\varepsilon}} = g(\phi)\boldsymbol{\sigma}^+ + \boldsymbol{\sigma}^- \tag{7}$$

where $\boldsymbol{\sigma}^\pm = \lambda\langle\text{tr}(\boldsymbol{\varepsilon})\rangle^\pm \mathbf{I} + 2\mu\boldsymbol{\varepsilon}^\pm$ are positive and negative stress tensors, which are proposed by Miehe [22] as an anisotropic phase-field approach.

The strong form equation is first introduced for the fourth-order phase-field CZM, resulting in a combination of the fourth-order formulation [36] and the non-standard phase-field CZM [50], given as:

$$\frac{\mathcal{G}_C}{c_\alpha}\left[\frac{\alpha'(\phi)}{l_0} - l_0\Delta\phi + \frac{l_0^3}{8}\Delta(\Delta\phi)\right] + g'(\phi)\mathcal{H}^+ = 0 \text{ on } \Omega, \tag{8}$$

with boundary conditions $\left[\nabla\phi - \frac{l_0^2}{8}\nabla(\Delta\phi)\right]\cdot\mathbf{n} = 0$ and $\Delta\phi = 0$ on $\partial\Omega$.

The first derivatives of the geometric function and energetic degradation function are $\alpha'(\phi) = \partial\alpha/\partial\phi$ and $g'(\phi) = \partial g/\partial\phi$, respectively. A history-field variable $\mathcal{H}^+ \coloneqq \max(\mathcal{H}_0, \mathcal{H}_n)$ which is proposed by Wu [61], is given as the maximum number of the initial damage driving force $\mathcal{H}_0 = f_t^2/(2E_0)$ and the damage driving force $\mathcal{H}_n = \sigma_{eq}^2/(2E_0)$ at time-step *n*. Parameters $E_0$ and $f_t$ are modulus of elasticity and yield strength parameters, respectively. For quasi-brittle materials, the equivalent effective stress $\sigma_{eq}$ with Rankine criterion is given as:

$$\sigma_{eq}(\boldsymbol{\sigma}^+) = \langle\sigma_1\rangle^+ \tag{9}$$

where $\sigma_1$ is the major principal stress component, which is computed from the positive stress tensor $\boldsymbol{\sigma}^+$.

### 2.3. Weak form equations

The variational principle is applied to derive to a weak form equation from Eq. (6), which is given as:

$$\int_\Omega \boldsymbol{\sigma}\delta\boldsymbol{\varepsilon}\mathrm{d}\Omega - \int_\Omega \mathbf{b}\cdot\delta\mathbf{u}\mathrm{d}\Omega - \int_{\partial\Gamma_t} \mathbf{t}\cdot\delta\mathbf{u}\mathrm{d}\Gamma_t = 0. \tag{10}$$

Additionally, the weak form equations of the phase-field variable for fourth-order phase-field CZM which are expressed as

$$\int_\Omega \left\{ \frac{\mathcal{G}_C}{c_\alpha} \left[ \frac{\alpha'(\phi)}{l_0}\delta\phi + l_0 \nabla\phi\cdot\nabla\delta\phi + \frac{l_0^3}{8}\Delta\phi\cdot\Delta\delta\phi \right] + g'(\phi)\mathcal{H}^+\delta\phi \right\} \mathrm{d}\Omega = 0 \tag{11}$$

### 2.4. Discretization

As described in Eq. (11), the fourth-order formulation for CZM includes a second derivative of the phase-field variable. Therefore, it requires that the approximation must fulfill at least $C^1$-continuity, which cannot be satisfied by the traditional finite element method. The best candidate, which can satisfy this requirement, is Isogeometric analysis (IGA) [53]. IGA has been demonstrated as an effective computational tool in several problems. This approach can construct the degree of smoothness very flexibly by choosing the interpolation order of basis functions, built on the Non-uniform rational B-Splines (NURBS).

The displacement and the phase-field approximations are given as

$$\mathbf{u} = \sum_{i=1}^n N_i^\mathbf{u} \mathbf{u}_i, \quad \phi = \sum_{i=1}^n N_i \phi_i \tag{12}$$

where $N_i$ is the NURBS basis function, and $n$ is the total number of control points per element. Particularity, the NURBS basis functions are described in very detail by Piegl [62]. The interpolation matrix, corresponding to the control point $i^{th}$ of the element in the domain, is defined as

$$N_i^\mathbf{u} = \begin{bmatrix} N_i & 0 \\ 0 & N_i \end{bmatrix}. \tag{13}$$

The strain tensor is expressed as

$$\boldsymbol{\varepsilon} = \sum_{i=1}^n \mathbf{B}_i^u \mathbf{u}_i \tag{14}$$

where the displacement-strain matrix is given as:

$$\mathbf{B}_i^\mathbf{u} = \begin{bmatrix} N_{i,x} & 0 & N_{i,y} \\ 0 & N_{i,y} & N_{i,x} \end{bmatrix}^T. \tag{15}$$

The gradient and Laplace's operator of the phase-field variable are given as:

$$\nabla \phi = \sum_{i=1}^{n} \mathbf{B}_{\mathbf{i}}^{\phi} \phi_i, \quad \Delta \phi = \sum_{i=1}^{n} D_i^{\phi} \phi_i \qquad (16)$$

where the first and second derivative terms of the phase-field are computed as follows:

$$\mathbf{B}_{\mathbf{i}}^{\phi} = \begin{bmatrix} N_{i,x} & N_{i,y} \end{bmatrix}^T, \quad D_i^{\phi} = N_{i,xx} + N_{i,yy}. \qquad (17)$$

The first and second derivatives of the shape function are represented by $N_{i,x}$, $N_{i,y}$ and $N_{i,xx}$, $N_{i,yy}$ concerning physical coordinations $x$ and $y$, respectively.

Similarly, the virtual variables and the derivatives are given:

$$\delta \mathbf{u} = \sum_{i=1}^{n} \mathbf{N}_i^{\mathbf{u}} \delta \mathbf{u}_i, \quad \delta \phi = \sum_{i=1}^{n} N_i \delta \phi_i, \quad \delta \epsilon = \sum_{i=1}^{n} \mathbf{B}_i^{\delta \mathbf{u}} \mathbf{u}_i, \quad \nabla \delta \phi = \sum_{i=1}^{n} \mathbf{B}_{\mathbf{i}}^{\phi} \delta \phi_i \text{ and } \Delta \delta \phi = \sum_{i=1}^{n} D_i^{\phi} \delta \phi_i. \qquad (18)$$

## 2.5. *A monolithic scheme*

The equilibrium equation, which can be derived as a weak form of external and internal work increment for the quasi-static problem, is expressed as:

$$\delta W_{ext} - \delta W_{int} = 0 \qquad (19)$$

where $\delta W_{ext}$ and $\delta W_{int}$ are the virtual works of external and internal energies which are shown in Eqs. (20) and (21).

$$\delta W_{ext} = \int_{\Omega} \mathbf{b} \cdot \delta \mathbf{u} d\Omega + \int_{\partial \Gamma_t} \mathbf{t} \cdot \delta \mathbf{u} d\Gamma_t \qquad (20)$$

$$\delta W_{int} = \int_{\Omega} \frac{\mathcal{G}_C}{c_\alpha l_0} \alpha'(\phi) \delta \phi d\Omega + \int_{\Omega} \frac{\mathcal{G}_C l_0}{c_\alpha} \nabla \phi \cdot \nabla \delta \phi d\Omega + \int_{\Omega} \frac{\mathcal{G}_C l_0^3}{8 c_\alpha} \Delta \phi \cdot \Delta \delta \phi d\Omega \\ + \int_{\Omega} g'(\phi) \mathcal{H}^+ \delta \phi d\Omega + \int_{\Omega} \boldsymbol{\sigma} \delta \boldsymbol{\epsilon} d\Omega \qquad (21)$$

As a result, the equilibrium equations are givens as

$$\int_{\Omega} \frac{\mathcal{G}_C}{c_\alpha l_0} \alpha'(\phi) \delta \phi d\Omega + \int_{\Omega} \frac{\mathcal{G}_C l_0}{c_\alpha} \nabla \phi \cdot \nabla \delta \phi d\Omega + \int_{\Omega} \frac{\mathcal{G}_C l_0^3}{8 c_\alpha} \Delta \phi \cdot \Delta \delta \phi d\Omega \\ + \int_{\Omega} g'(\phi) \mathcal{H}^+ \delta \phi d\Omega + \int_{\Omega} \boldsymbol{\sigma} \delta \boldsymbol{\epsilon} d\Omega - \int_{\Omega} \mathbf{b} \cdot \delta \mathbf{u} d\Omega - \int_{\partial \Gamma_t} \mathbf{t} \cdot \delta \mathbf{u} d\Gamma_t = 0 \qquad (22)$$

To solve the non-linear equation by minimizing the internal potential energy, Miehe [22] first has proposed a monolithic scheme for the crack propagation problem as a coupled-field problem using a displacement control algorithm of the Newton-Raphson iteration. This scheme allows obtaining either displacement and phase-field variables in each iteration. A flow chart of the monolithic algorithm is illustrated in Figure 2.

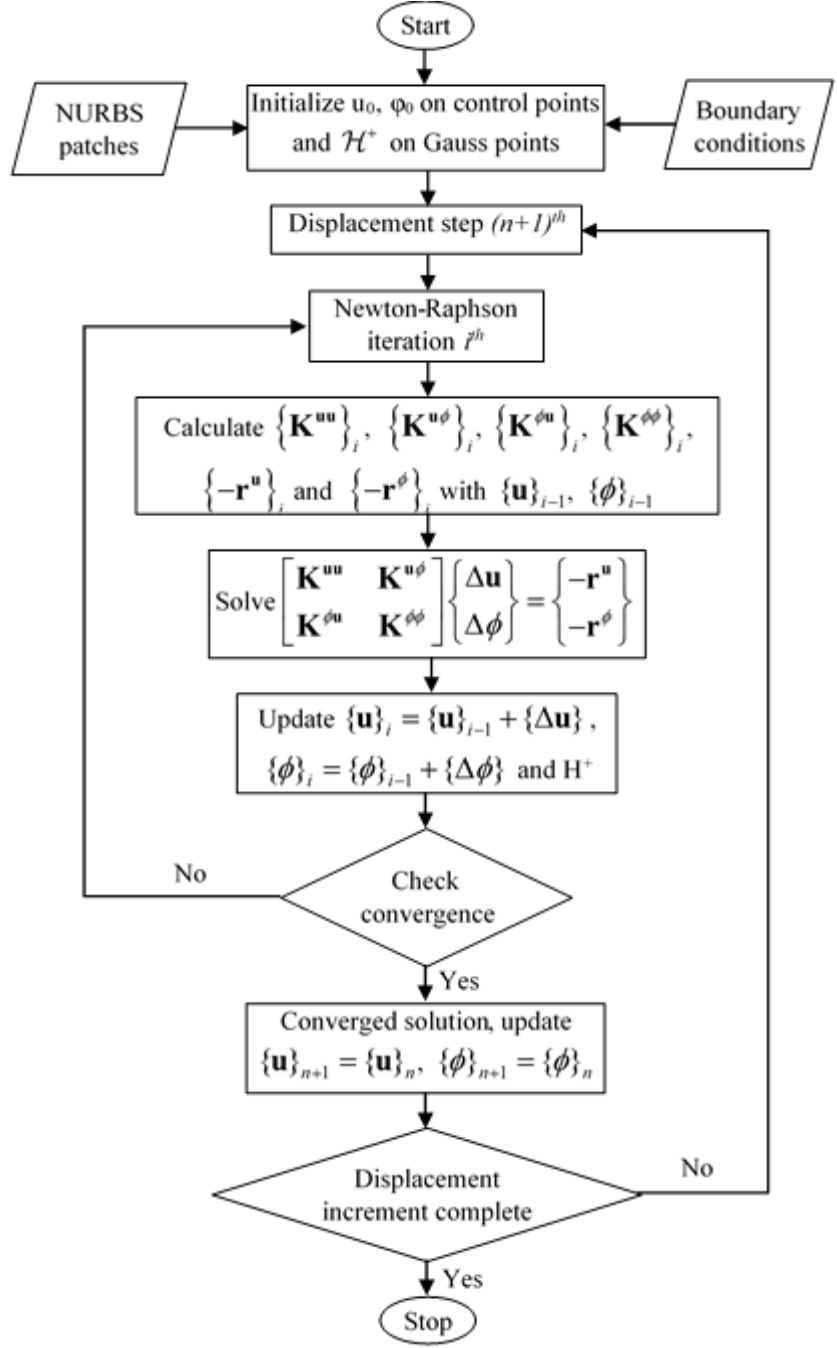

Figure 2. A flow chart algorithm for the monolithic scheme.

In this flow chart, the residual force vector $\mathbf{r}_i^u$ of the displacement field and the residual force vector $\mathbf{r}_i^\phi$ of the phase-field variable at $i^{th}$ iteration is given as:

$$\mathbf{r}_i^u = \int_\Omega (\mathbf{B}_i^u)^T \boldsymbol{\sigma} d\Omega - \int_\Omega (\mathbf{N}_i^u)^T \mathbf{b} d\Omega - \int_{\partial \Gamma_t} (\mathbf{N}_i^u)^T \mathbf{t} d\partial\Gamma_t \qquad (23)$$

$$r_i^\phi = \int_\Omega \left\{ \frac{\mathcal{G}_C}{c_\alpha} \left[ \frac{\alpha'}{l_0} N_i + l_0 (\mathbf{B}_i^\phi)^T \nabla\phi + \frac{l_0^3}{8} D_i^\phi \Delta\phi \right] + g'\mathcal{H}^+ N_i \right\} d\Omega. \qquad (24)$$

From the residual force vectors, the tangent stiffness matrices can be computed by Eqs. (25)-(27).

$$\mathbf{K}_{ij}^{\mathbf{uu}} = \frac{\partial \mathbf{r}_i^{\mathbf{u}}}{\partial \mathbf{u}_j} = \int_\Omega (\mathbf{B}_i^{\mathbf{u}})^T \mathbf{C} \mathbf{B}_j^{\mathbf{u}} d\Omega, \tag{25}$$

$$\mathbf{K}_{ij}^{\mathbf{u}\phi} = \frac{\partial \mathbf{r}_i^{\mathbf{u}}}{\partial \phi_j} = \int_\Omega (\mathbf{B}_i^{\mathbf{u}})^T g' \boldsymbol{\sigma}^+ N_j d\Omega, \quad \mathbf{K}_{ij}^{\phi\mathbf{u}} = \frac{\partial r_i^\phi}{\partial \mathbf{u}_j} = \int_\Omega g' N_i \frac{\partial \mathcal{H}^+}{\partial \boldsymbol{\varepsilon}} \mathbf{B}_i^{\mathbf{u}} d\Omega, \text{ and} \tag{26}$$

$$\mathbf{K}_{ij}^{\phi\phi} = \frac{\partial r_i^\phi}{\partial \phi_j} = \int_\Omega \left\{ \frac{\mathcal{G}_C l_0}{c_\alpha} (\mathbf{B}_i^\phi)^T \mathbf{B}_j^\phi + \left[ g'' \mathcal{H}^+ + \alpha'' \frac{\mathcal{G}_C}{c_\alpha l_0} \right] N_i N_j + \frac{\mathcal{G}_C l_0^3}{8 c_\alpha} D_i^\phi D_j^\phi \right\} d\Omega. \tag{27}$$

In Eq. (25), the fourth-order tensor $\mathbf{C}$ is defined in detail in the previous work [57].

*2.6. Characteristic functions*

Two functions of the geometric crack function $\alpha(\phi)$ and the energetic degradation function $g(\phi)$ characterize the phase-field CZM. The former regularizes the sharp crack topology while the latter defines the free energy density potential. Particularity, the characteristic functions of the standard phase-field model, proposed by Miehe [22], is used to compute damage for the brittle failure, given as

$$\alpha(\phi) = \phi^2 \Rightarrow \alpha' = 2\phi, \ \alpha'' = 2, \ c_\alpha = 2, \tag{28}$$

$$g(\phi) = (1-\phi)^2 \Rightarrow g' = -2(1-\phi), \ g'' = 2. \tag{29}$$

It is easy to see that the standard formulations can be recovered by substituting Eqs. (28) and (29) into Eqs. (23)-(27).

In addition, Wu [50, 61] proposed these characteristic functions to compute damage for the quasi-brittle failure with a cohesive zone model, expressed as

$$\alpha(\phi) = 2\phi - \phi^2 \Rightarrow \alpha' = 2 - 2\phi, \ \alpha'' = -2, \ c_\alpha = \pi, \tag{30}$$

$$g(\phi) = \frac{(1-\phi)^n}{(1-\phi)^n + a_1 \phi (1 + a_2 \phi + a_3 \phi^2)}. \tag{31}$$

The first parameter is chosen as $a_1 = 4 l_{ch} / (\pi l_0)$ where $l_{ch} = E_0 \mathcal{G}_C / f_t^2$ is Irwin's characteristic length of material. Next, the optimal model parameters [61] are chosen by choice of the softening laws, which are given as:

$$\text{Linear softening curve: } n = 2.0, \ a_2 = -0.5, \ a_3 = 0.0, \tag{32}$$

$$\text{Exponential softening curve: } n = 2.5, \ a_2 = 0.1748, \ a_3 = 0.0, \tag{33}$$

$$\text{Hyperbolic softening curve: } n = 4.0, \ a_2 = 0.5397, \ a_3 = 0.0, \tag{34}$$

$$\text{Cornelissen softening curve: } n = 2.0, \ a_2 = 1.3868, \ a_3 = 0.9106. \tag{35}$$

This study focuses on the damaged problem on concrete material, so the parameters are chosen with Cornelissen softening curve for all numerical results. Also, it is realized that the energetic degradation function in Eq. (31) becomes Eq. (29) as the standard phase-field model by applying the parameters of $n = 2.0$, $a_1 = 2.0$, $a_2 = -0.5$, $a_3 = 0.0$.

3. **Numerical examples**

This section introduces several numerical examples to demonstrate the efficiency of the fourth-order CZM formulation in computing damage on the quasi-brittle material. Also, the approach includes the second derivative, so it requires at least the $C^1$-continuity ability of the approximation method. IGA shows as an outstanding candidate in this study because it is based on NURBS basis functions to choose appropriate interpolation order of basis functions. Because the tensor-product topology generates NURBS basis functions, they cause difficulty building free-form geometries using multi-patch mesh. Therefore, a refinement of multi-patch mesh usually is a global refinement mesh. Besides, the size of the elements, using the phase-field formulation, is required to be small enough to represent a crack pattern topology accurately. Typically, the global refinement mesh of the traditional IGA will cause many elements that consume a large amount of memory and computational time. Hence, the VUKIMS coupling algorithm [58] was developed successfully to overcome this issue for the phase-field approach in the previous study [57]. It helps to build a local refinement mesh for the multi-patch model. Here, the minimum degree of approximation should be chosen at least quadratic to fulfill the $C^1$ condition of the fourth-order formulation.

Here, the non-linear equations are solved by the displacement control algorithm of the Newton-Raphson iteration. The criterion of the convergence, which is based on the residual force vectors, is given as

$$max(\|\mathbf{r}_i^\phi\|, \|\mathbf{r}_i^\mathbf{u}\|) \leq \mathcal{TOL} = 1 \times 10^{-4} \tag{36}$$

where $\mathbf{r}_i^\phi$ and $\mathbf{r}_i^\mathbf{u}$ are the residual force vectors of the phase-field and displacement fields, as given in Eqs. (23) and (24).

*3.1. Asymmetric three-point bending concrete beam*

The first example is an asymmetric three-point bending beam conducted by Rots [63]. The beam is composed of concrete material, which is assumed as a quasi-brittle material. The geometry, boundary conditions, and loading are also shown in Figure 3. The specimen is considered as in plane-stress condition with a thickness of 100 mm. An expectation of the crack propagation is vertically from the bottom to the top edge because a single notch tip of the beam is located on the line of symmetry. The material properties chosen from Rots's study are Young's modulus $E = 20 \text{ GPa}$, Poisson's ratio $v = 0.2$, the critical energy release density $\mathcal{G}_C = 0.000113 \text{ kN/mm}$, and the yield strength $f_t = 0.0024 \text{ GPa}$. As in the previous study [50], a length-scale number $l_0 = 2.5 \text{ mm}$ is chosen.

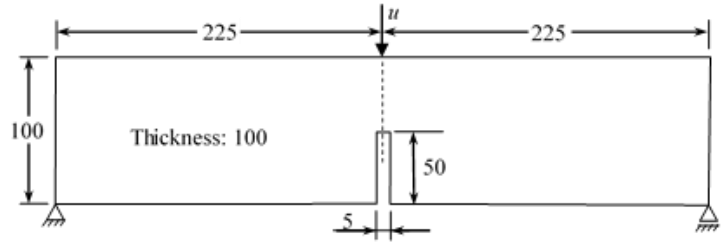

Figure 3. Asymmetric three-point bending concrete beam (unit of length: mm).

An incremental monotonic displacement $\Delta u = 0.01$ mm is subjected to the center point of the upper edge for the 100 loading steps. With the effective size $h = l_0/2$, Figure 4 illustrates the local refinement mesh of the concrete beam using the VUKIMS coupling algorithm to reduce the required memory compared to the global refinement mesh of the traditional IGA. The advantages of VUKIMS have been demonstrated in the previous work [57]. In this study, the fourth-order CZM formulation is investigated to compute crack propagation in the quasi-brittle material. This model requires at least $C^1$-continuity ability, so the order approximation of B-spline elements is chosen at least second-order (quadratic elements). Table 1 shows the DOF numbers and the computational time corresponding to each mesh case of B-spline elements. A computer using AMD Ryzen 7 2700X processor is used to evaluate all of the time-consuming tests. Four different effective element sizes $h = 2l_0, l_0, l_0/2$, and $l_0/4$, corresponding to three order approximations (quadratic, cubic, and quartic), are considered to evaluate a converged result (please see Figure 5a).

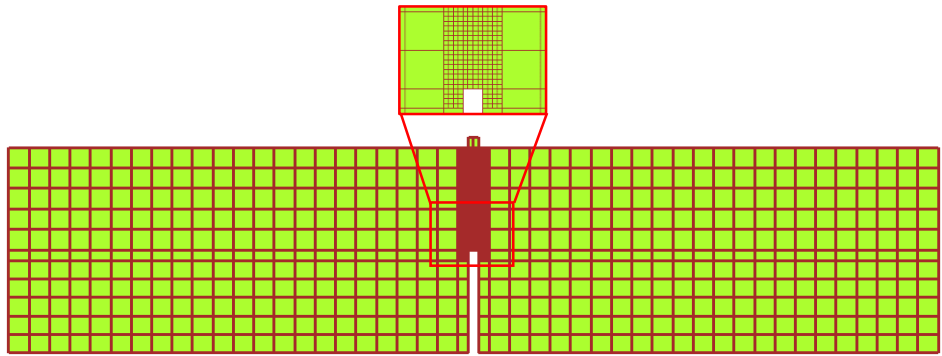

Figure 4. The local refinement mesh of symmetric three-point bending concrete beam.

As a result, Figure 5a demonstrates that almost all mesh cases are in good agreement compared to Rots's study's experimental result [63]. Also, the predicted crack patterns of various meshes, as illustrated in Table 2, are confirmed very well with the numerical result using the second-order phase-field CZM from Wu's work [50]. The crack paths and the peak load results are insensitive to all mesh cases. Although quartic elements help us gain outstanding results in all effective element sizes, their performance is worse than the cubic ones because the formers consume over double computational time in the same element size. Hence, to balance both the accuracy and computational cost of the solutions, the effective size $h = l_0 / 2$ of the cubic elements seems to be appropriate for the proposed approach.

Furthermore, Figure 5b shows reaction force versus displacement curves in two cases of the second-order and fourth-order phase-field CZM, using the cubic elements with the effective element size $h = l_0 / 2$. These results are compared with either the experimental and numerical results [50, 63]. They reveal that the curve of the second-order CZM approach is confirmed very well with the result from [50] using the same model, while the fourth-order phase-field CZM result is in good agreement in comparison with [63]. Additionally, the latter performs more accurately because its curve is more fit in the range of the experimental result than the former. Hence, the effective size $h = l_0 / 2$ of the cubic element mesh is recommended to use in the following problems.

Table 1

DOF number and computational time of the multiple meshes of B-spline elements.

| Order approximation | Effective element size $h$ | DOF number | Computational time (min) |
|---|---|---|---|
| quadratic | $2*l_0$ | 2238 | 3 |
| quadratic | $l_0$ | 2772 | 4 |
| quadratic | $l_0/2$ | 4488 | 9 |
| quadratic | $l_0/4$ | 10224 | 27 |
| cubic | $2*l_0$ | 2912 | 6 |
| cubic | $l_0$ | 3584 | 9 |
| cubic | $l_0/2$ | 5582 | 23 |
| cubic | $l_0/4$ | 11882 | 70 |
| quartic | $2*l_0$ | 3662 | 16 |
| quartic | $l_0$ | 4472 | 24 |
| quartic | $l_0/2$ | 6752 | 50 |
| quartic | $l_0/4$ | 13616 | 172 |

Table 2

Damage profile of the multiple meshes of B-spline elements.

| Order approximation | Effective size h | | | |
|---|---|---|---|---|
| | $2*l_0$ | $l_0$ | $l_0/2$ | $l_0/4$ |
| quadratic | | | | |
| cubic | | | | |
| quartic | | | | |

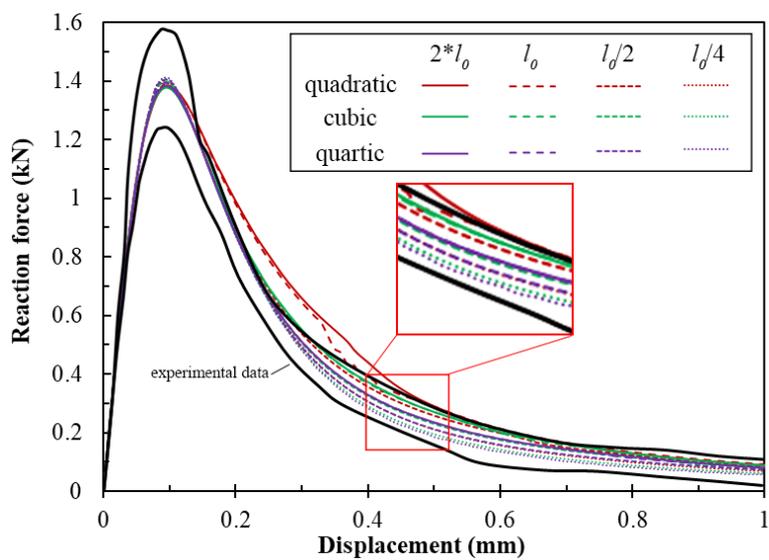

(a)

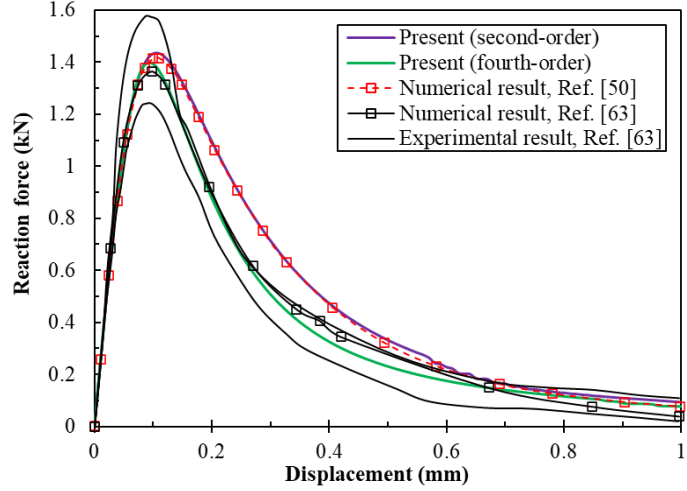

(b)

Figure 5. Reaction force versus displacement curves of the concrete beam.

*3.2.  L-shaped panel test*

The second example considers a concrete L-shaped panel test as a well-known benchmark problem of the mixed-mode failure for the concrete structure [64-67], conducted by Winkler [68]. The stress state is assumed as a plane-stress condition with a thickness of 100 mm. The geometry and boundary conditions of an L-shaped panel are displayed in Figure 6a, while Figure 6b describes the cubic B-Spline element mesh with the effective element size $h = l_0/2$, where a length-scale parameter is chosen $l_0 = 7.5$ mm. Hence, the mesh has approximately 7042DOFs. The material parameters are chosen from [67] as Young's modulus $E = 20$ GPa, Poisson's ratio $\nu = 0.18$, the yield strength $f_t = 0.0025$ GPa, and the critical energy release density $\mathcal{G}_C = 0.00013$ kN/mm.

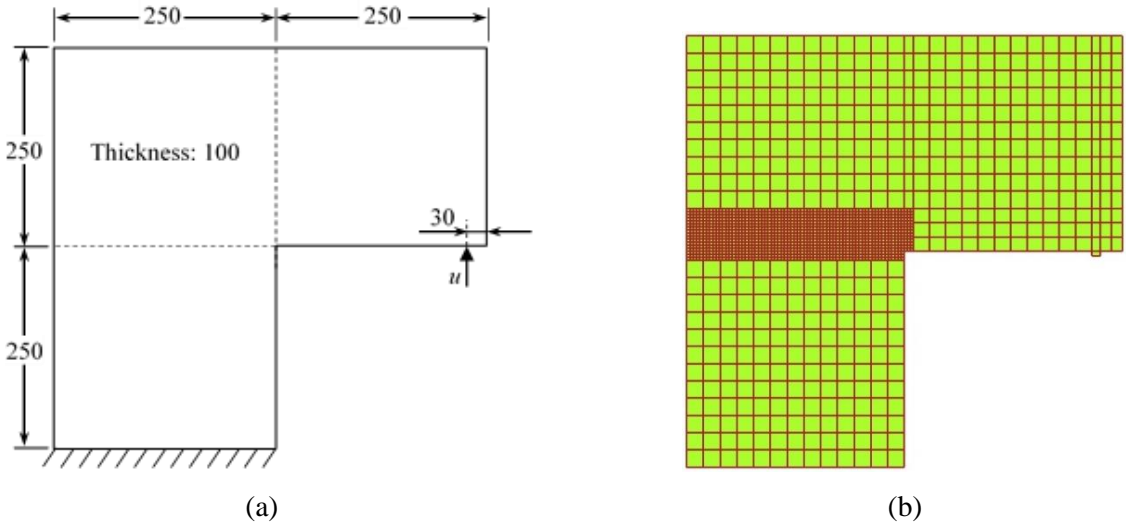

(a)                                                             (b)

Figure 6. L-shaped panel test: (a) geometry (unit of length: mm), boundary conditions, and (b) local refinement mesh.

Several studies used the extended finite element method [64-67], which requires an initial crack surface. Also, the orientation of the pre-existing crack will cause a different crack path. This issue can be overcome without any initialization of the crack when using the phase-field model. Wu [50, 61] has applied the second-order CZM formulation with a staggered scheme to predict crack propagation for this problem. We propose the fourth-order CZM formulation to combine the anisotropic fourth-order formulation [36] and the cohesive zone model [50]. In this problem, the vertical displacement increment $\Delta u = 0.02$ mm is applied to the point located at a distance of 30 mm to the right edge (please see Figure 6a). The numerical predictions of the L-shaped panel test's reaction force versus displacement curves, as illustrated in Figure 7, are in good agreement in all cases of the length-scale parameter with the experimental results from [69]. Significantly, the current results are more accurate than the numerical result using the second-order model [61] compared to the experimental results when the displacement is over 0.6 mm.

Additionally, Figure 8 shows the predicted crack paths which use the fourth-order phase-field CZM without any initialization of the pre-existing crack with four different length-scale parameters ($l_0 = 2.5,\ 5.0,\ 7.5$ and $10.0$ mm). It reveals that the predicted crack paths and the peak load results of the fourth-order phase-field CZM approach are not merely a good confirmation compared with the experimental ranges (described as a grey area) taking from Winkler [68] and the numerical solutions from several pieces of literature [50, 61, 64-67], but also independent of the incorporated length-scale number. Therefore, the current approach has a significant benefit to apply to the three-dimensional failure problem because it can minimize the computational cost with an appropriate selection of the length-scale parameter.

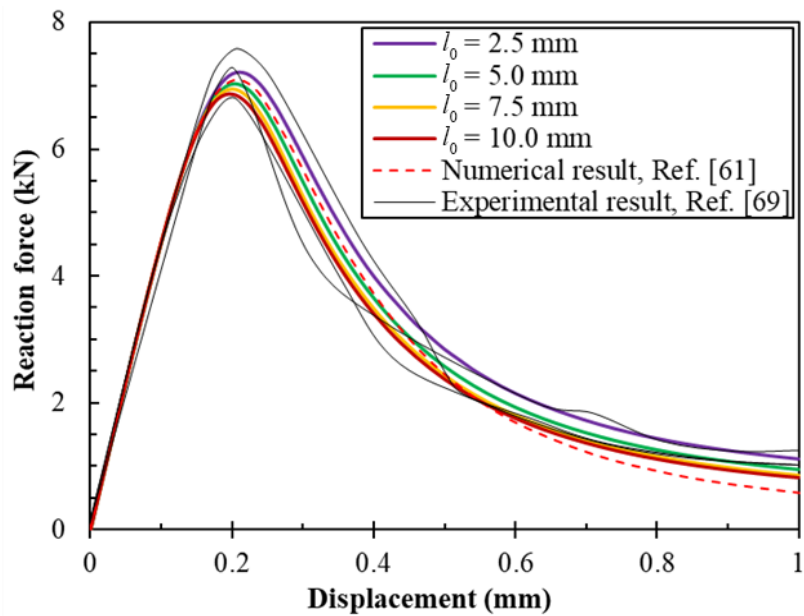

Figure 7. Reaction force versus displacement curves of the L-shaped panel test.

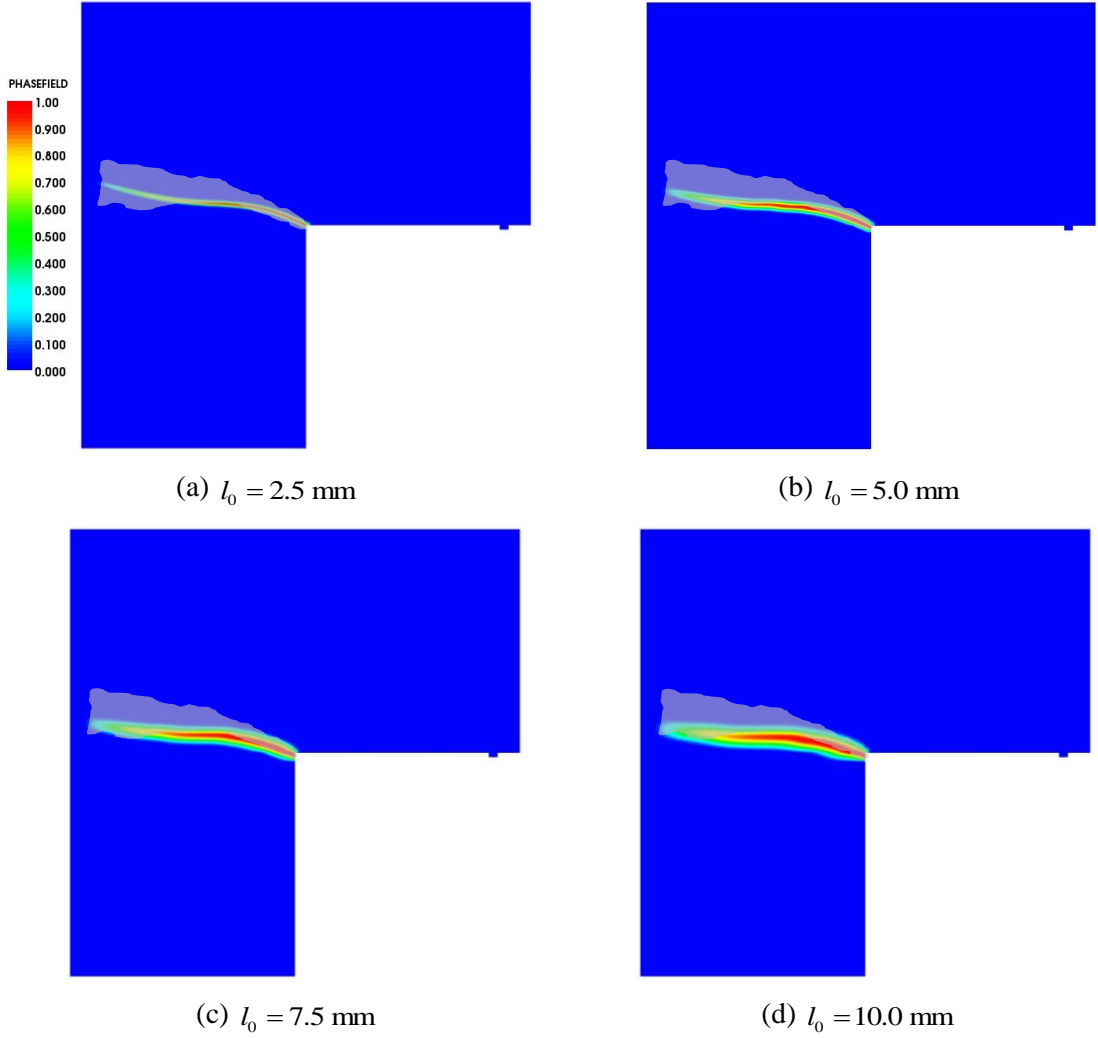

(a) $l_0 = 2.5$ mm

(b) $l_0 = 5.0$ mm

(c) $l_0 = 7.5$ mm

(d) $l_0 = 10.0$ mm

Figure 8. The L-shaped panel test predicted crack path compared with an experimental result from Winkler [68].

### 3.3. *Mixed-mode failure of three-point bending concrete beam*

The last example considers a mixed-mode failure of the three-point bending concrete beam with its geometry and boundary conditions, as reported in [70, 71], which are illustrated in Figure 9. The displacement increment $\Delta u = 0.001$ mm is subjected to the middle point of the beam until the crack mouth opening displacement (CMOD) equals 0.6 mm. The stress state is assumed as a plane-stress condition with a constant thickness of 50 mm. Here, the beam's three heights ($H$) are 80 mm, 160 mm, and 320 mm. The length ratio of the height, as shown in Figure 9, is considered as three cases of $a = 0$, 0.3125, and 0.625. The concrete material properties from [70] are as follows: Young's modulus $E = 33.8$ GPa, Poisson's ratio $\nu = 0.2$, the yield strength $f_t = 0.0035$ GPa, and the critical energy release density $\mathcal{G}_C = 0.00008$ kN/mm. Depending on the height ($H$) of the beam, a length-scale number is chosen as $l_0 = 0.001H$ (mm). The fourth-order CZM formulation is only considered in this example. Also, the effective size $h = l_0 / 2$ of the cubic elements is utilized for the predicted crack propagation zones.

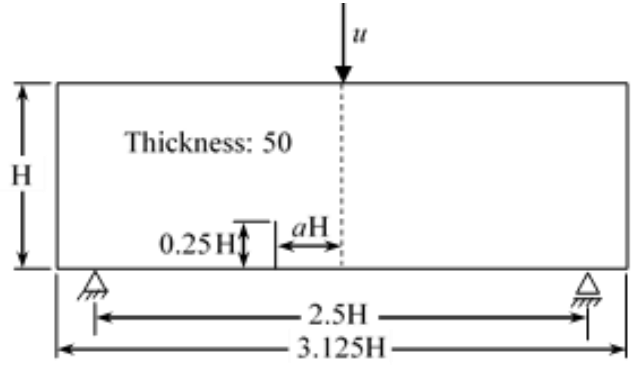

Figure 9. The geometry and boundary conditions of mixed-mode concrete beam test (unit of length: mm).

As can be seen, the reaction force curves versus CMOD for the mixed-mode concrete beam are illustrated in Figure 10. As a result, predictions of the peak loads are in excellent agreement compared with the experimental results from [70], as well as the numerical results using the re-mesh technique [70] from and the second-order phase-field CZM approach from [71]. Also, Figure 11 shows that the predicted crack patterns, using the fourth-order phase-field CZM, have been confirmed for three-point bending concrete beam within the experimental ranges (described as grey areas) from [70] in most cases. It can be easily observed that the initial crack position, located more near the center of the beam, will cause a lower peak load as the same height of the beam.

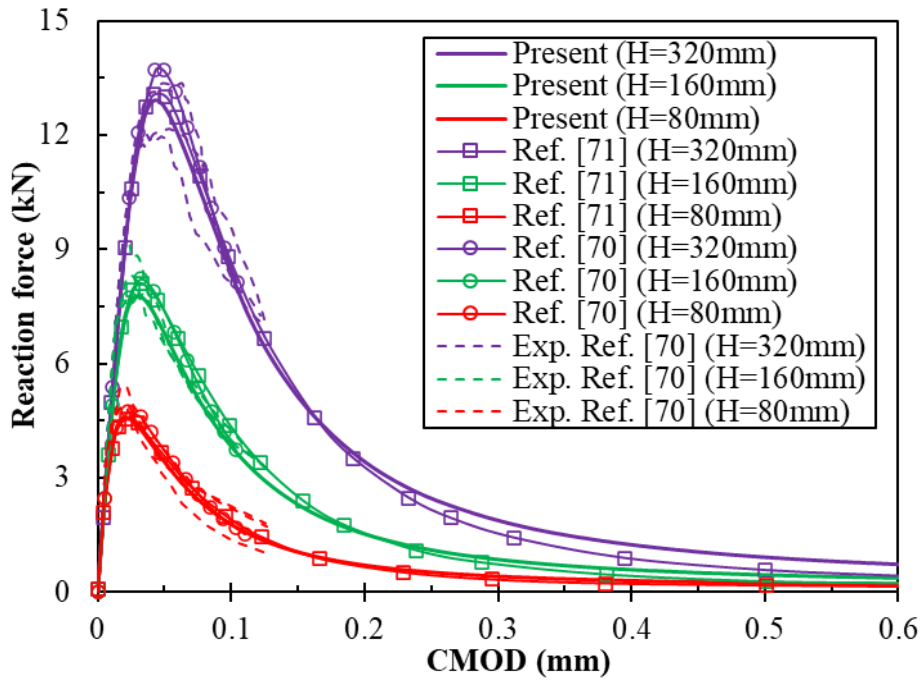

(a) $a = 0.6250$

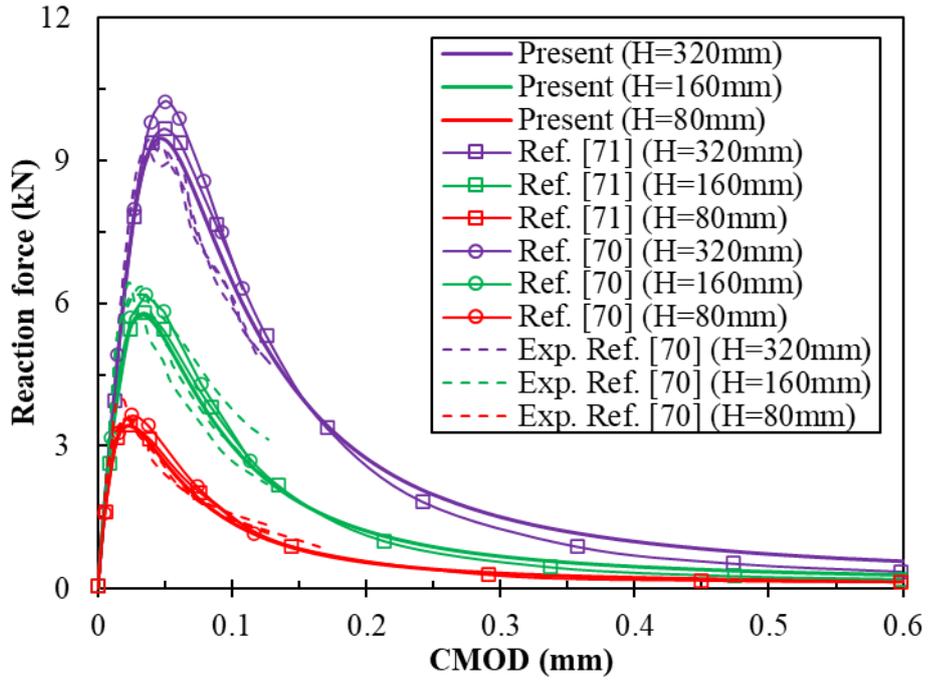

(b) $a = 0.3125$

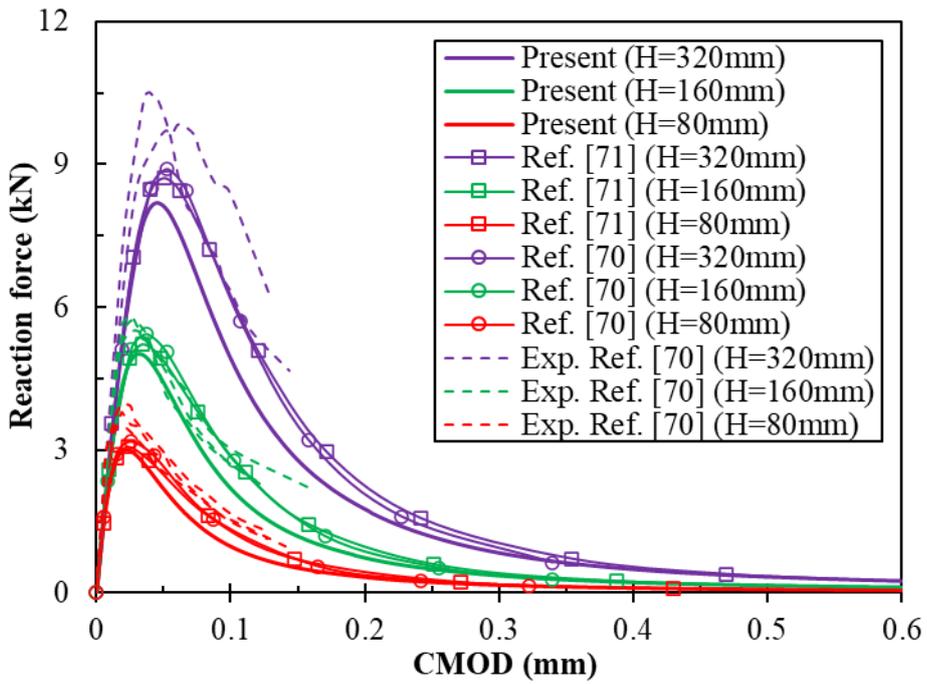

(c) $a = 0$

Figure 10. Reaction force versus CMOD curves of the mixed-mode concrete beam.

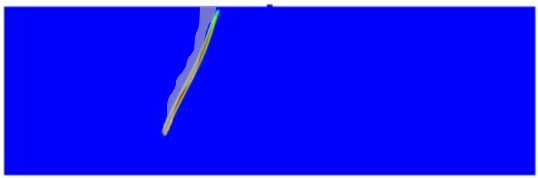
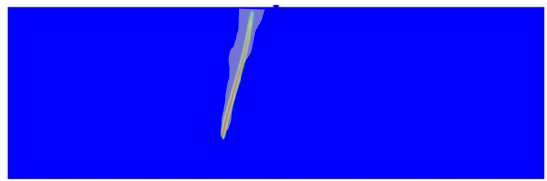

(a) H=320mm, a=0.6250          (d) H=320mm, a=0.3125

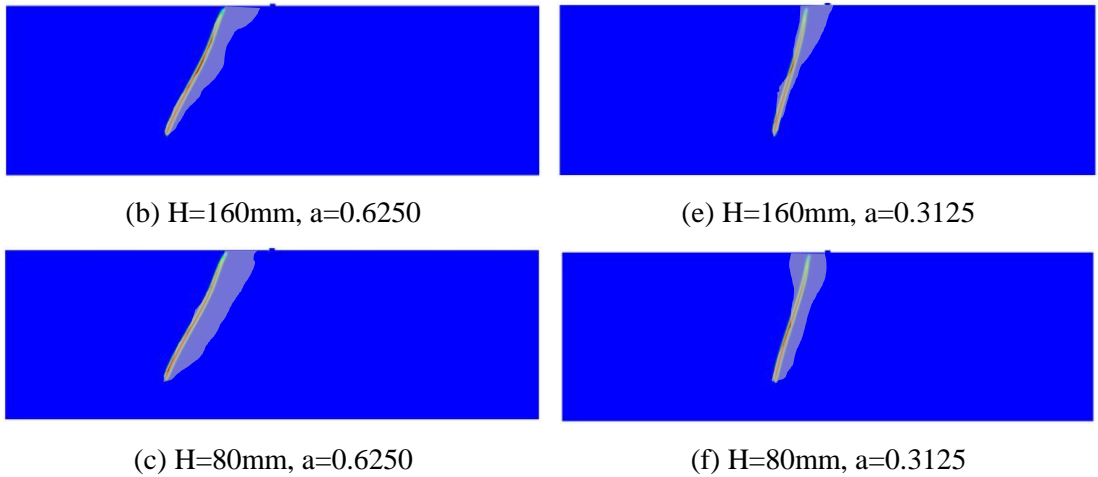

(b) H=160mm, a=0.6250    (e) H=160mm, a=0.3125

(c) H=80mm, a=0.6250    (f) H=80mm, a=0.3125

Figure 11. The predicted crack patterns of the mixed-mode concrete beam compared with an experimental result from Ref. [70].

## 4. Conclusion

In this study, the fourth-order phase-field CZM, as a non-standard phase-field model, was investigated for the quasi-brittle fracture. The numerical results were verified with the proposed ones to demonstrate the accuracy of the approach. The effective size $h = l_0/2$ of the cubic NURBS element mesh was recommended to archive the accurate solutions. As a result, the proposed approach improved the accuracy of the predicted solution compared with the second-order one [50]. In addition, the solutions revealed that both the crack path and the peak load were insensitive to the length-scale number and independent of the element size. These characteristics help us choose an appropriate length-scale number and the element size to minimize computational cost and guarantee the required accuracy for the subsequent studies. Furthermore, because of the benefits of the fourth-order phase-field CZM model, the current approach is a promising method in applying to the complex practical computations, including more complicated geometries on three-dimensional structures involving curved boundaries and more general softening laws by using different optimal characteristic functions. They would be considered in the subsequent research.

## Acknowledgments

The authors acknowledge VLIR-UOS TEAM Project's financial support, VN2017TEA454A103, "An innovative solution to protect Vietnamese coastal riverbanks from floods and erosion", funded by the Flemish Government. In addition, the support provided by RISE-project BESTOFRAC (734370)–H2020 is gratefully acknowledged.